\documentclass[a4paper]{PoS}

\title{Determining the QCD coupling from lattice vacuum polarization}

\ShortTitle{Determining the QCD coupling from lattice vacuum polarization}

\author{Renwick J.\ Hudspith,$^a$ \speaker{Randy Lewis},$^a$ Kim Maltman$^{b,c}$ and Eigo Shintani$^d$ \\
        \llap{$^a$} Department of Physics and Astronomy, York University, Toronto, ON, M3J 1P3, Canada \\
        \llap{$^b$} Department of Mathematics and Statistics, York University, Toronto, ON, M3J 1P3, Canada \\
        \llap{$^c$} CSSM, University of Adelaide, Adelaide, SA, 5005, Australia \\
        \llap{$^d$} RIKEN Advanced Institute for Computational Science, Kobe, Hyogo 650-0047, Japan \\
        E-mail: \email{renwick.james.hudspith@googlemail.com}, \email{randy.lewis@yorku.ca}, \email{kmaltman@yorku.ca}, \email{shintani@riken.jp}}

\abstract{The QCD coupling appears in the perturbative expansion of the current-current two-point (vacuum polarization) function. Any lattice calculation of vacuum polarization is plagued by several competing non-perturbative effects at small momenta and by discretization errors at large momenta.  We work in an intermediate region, computing the vacuum polarization for many off-axis momentum directions on the lattice.  Having many momentum directions provides a way to monitor and account for lattice artifacts.  Our results are competitive with, and have certain systematic advantages over, the alternate phenomenological determination of the strong coupling from the same light quark vacuum polarization produced by sum rule analyses of hadronic $\tau$ decay data.}

\FullConference{The 33rd International Symposium on Lattice Field Theory\\
		 14 -18 July  2015\\
		 Kobe International Conference Center, Kobe, Japan}

\begin{document}

\section{Motivation}

$\alpha_s$ is a fundamental parameter of QCD and
its numerical value is crucial input for most practical calculations in particle physics.
Interpretation of experimental data requires the value of $\alpha_s$ to handle QCD backgrounds.
Several lattice QCD methods have been employed to determine $\alpha_s$, including
the short-distance QCD potential,
Wilson loops,
the Schr\"odinger functional,
the ghost-gluon vertex,
current two-point functions with heavy valence quarks,
and vacuum polarization at short distances.
For a review and references, see \cite{Aoki:2013ldr}.

Our focus for the present work is the method of vacuum polarization at short distances,
studied as a function of Euclidean $Q^2$,
as pioneered by Shintani and collaborators in a sequence of papers \cite{Shintani:2008ga,Shintani:2010ph,Shintani:2014rta}.
Related studies can be found in \cite{Herdoiza:2014jta,Francis:2014yga}.
The perturbative expression is a function of $\alpha_s$.
At small $Q^2$, there are important non-perturbative (NP) contributions in addition to perturbation theory.
At large $Q^2$, there are important lattice artifacts in addition to perturbation theory.

Previous research \cite{Shintani:2008ga,Shintani:2010ph,Shintani:2014rta} used $Q^2$ as low as about 1 GeV$^2$
and included NP contributions in the fit via the operator product expansion (OPE).
This is problematic if successive OPE terms have comparable sizes with alternating signs and,
perhaps surprisingly, this phenomenon really can occur, as illustrated by sum rule fit results for the light quark $V$+$A$ polarization: \cite{Boito:2014sta}
\begin{eqnarray}
\Pi_{\rm OPE}^{(1+0)}(Q^2) &=& \sum_{k=0}^\infty\frac{C_{2k}}{Q^{2k}} \,, \\
C_{4,V+A} &=& +0.00268 {\rm ~GeV^4} \,, \\
C_{6,V+A} &=& -0.0125 {\rm ~GeV^6} \,, \\
C_{8,V+A} &=& +0.0349 {\rm ~GeV^8} \,, \\
C_{10,V+A} &=& -0.0832 {\rm ~GeV^{10}} \,, \\
C_{12,V+A} &=& +0.161 {\rm ~GeV^{12}} \,, \\
C_{14,V+A} &=& -0.191 {\rm ~GeV^{14}} \,, \\
C_{16,V+A} &=& -0.233 {\rm ~GeV^{16}} \,.
\end{eqnarray}
For numerical values of $C_{6,V}$ and $C_{8,V}$, see Table III of \cite{Boito:2014sta}.
This sequence of coefficients only produces a manageable OPE series for 
sufficiently large $Q^2$. With this potential danger in mind, we restrict our attention to $Q^2$ values large enough that all NP OPE terms are negligible in the present analysis.

\section{Method}

The vector current two-point correlation function with $I$=1 and $m_u$=$m_d$ is
\begin{equation}
\langle V_\mu V_\nu\rangle \equiv \Pi_{\mu\nu}(Q)
 = (Q^2\delta_{\mu\nu}-Q_\mu Q_\nu)\Pi(Q^2) \,.
\end{equation}
The QCD coupling will be obtained from $\Pi(Q^2)$.
There is a close relation to the $\tau$ decay determination of $\alpha_s$,
which uses experimental spectral data and finite-energy sum rule (FESR) analysis of the same $\Pi(Q^2)$,
but here we have certain systematic advantages.
In particular, our lattice calculation is performed directly at Euclidean $Q^2$ whereas the FESR approach \cite{Boito:2014yfa} requires calculation on a circle in the complex $Q^2$ plane that comes infinitesimally close to the Minkowski axis. The OPE is not a good description of Minkowski physics, so the FESR approach relies on certain weights that are chosen to minimize the impact of physics near the Minkowski axis.  The present lattice approach avoids the issue entirely by working exclusively with Euclidean $Q^2$.

The perturbative expression up to 6 loops, in the $\overline{\rm MS}$ renormalization scheme at scale $\mu$, is
\begin{equation}
\Pi(Q^2) = C - \frac{1}{4\pi^2}\left(t+\sum_{k=1}^5\left(\frac{\alpha_s(\mu)}{\pi}\right)^k\sum_{m=0}^{k-1}c^A_{km}\frac{t^{m+1}}{m+1}\right)
\end{equation}
where $C$ is a constant and $t=\ln(Q^2/\mu^2)$.
All coefficients are known except $c^A_{50}$, which has been estimated \cite{Baikov:2008jh}.

Discretization errors grow as $Q^2$ increases, but they can be managed.
Choosing the momentum to be along a single lattice axis is particularly undesirable, so
we have generated $\Pi(Q^2)$ for all possible momenta $Q_\mu$.
With those data in hand, we define $\hat v=(1,1,1,1)/2$ and calculate
\begin{equation}
(Q_\perp)_\mu = Q_\mu - (Q\cdot\hat v)\hat v_\mu \,.
\end{equation}
If $Q_\mu$ points along a lattice diagonal then $Q_\perp=0$.
For fixed $|Q^2|$, those $Q_\mu$ options aligned most closely with $\hat v_\mu$ have the smallest lattice artifacts so
we implement a maximum radius $|Q_\perp|_{\rm max}$.

We handle O(4)-breaking lattice artifacts via reflection averaging:
\begin{equation}\label{ReflAve}
\Pi_{\rm lat}(Q^2) = \frac{1}{12}\sum_{\mu=x,y,z,t}\sum_{\nu\neq\mu}\left(\frac{\Pi_{\mu\nu}(Q)-\Pi_{\mu\nu}(R_\mu Q)}{2Q_\mu Q_\nu}\right)
\end{equation}
where $R_\mu$ is a reflection operator in the $\mu$ direction.
In practice, division by zero is no problem because any $Q$ having a vanishing component will be beyond the maximum radius $|Q_\perp|_{\rm max}$.
After reflection averaging, there are still O(4)-preserving artifacts that remain to be fitted:
\begin{equation}\label{fitfunction}
\Pi_{\rm lat}(Q^2) = \Pi(Q^2) + c_1a^2Q^2 + c_2a^4Q^4 + \ldots
\end{equation}

We use ensembles from the RBC and UKQCD Collaborations \cite{Arthur:2012yc} with parameters shown in Table~\ref{RBCUKQCD}.
\begin{table}
\caption{The lattices used in this study are characterized by spatial extent $L$, temporal extent $T$,
gauge coupling $\beta$, bare quark masses $m_s$ and $m_\ell$, lattice spacing $a$, vector renormalization
constant $Z_V$, and number of configurations $n_{\rm conf}$.}\label{RBCUKQCD}
\begin{center}
\begin{tabular}{ccccccc}
\hline\hline
$L^3\times T$ & $\beta$ & $m_s$ & $m_\ell$ & $a^{-1}$ [GeV] & $Z_V$ & $n_{\rm conf}$ \\
\hline
24$^3\times$64 & 2.13 & 0.04 & 0.005,0.01,0.02 & 1.78 & 0.714 & 900 \\
32$^3\times$64 & 2.25 & 0.03 & 0.004,0.006,0.008 & 2.38 & 0.745 & 940 \\
\hline\hline
\end{tabular}
\end{center}
\vspace{-5mm}
\end{table}
We calculate $\langle V_\mu^LV_\nu^C\rangle$.
The local current $V^L$ removes a contact term and
the conserved current $V^C$ preserves the Ward-Takahashi identity:
\begin{equation}
\sum_\nu\hat Q_\nu e^{iQ_\nu/2}\Pi_{\mu\nu} = 0
{\rm ~~where~~}
\hat Q_\nu = 2\sin(Q_\nu/2) \,.
\end{equation}
Because $\Pi(Q^2)$ only depends on $\alpha_s$ at subleading orders,
we prefer to use instead a renormalization-independent function where $\alpha_s$ appears at leading order:
\begin{eqnarray}
\Delta(Q^2,Q_{\rm ref}^2)
&\equiv& -4\pi^2\left(\frac{\Pi_{\rm lat}(Q^2)-\Pi_{\rm lat}(Q_{\rm ref}^2)}{\ln(\hat Q^2/\hat Q_{\rm ref}^2)}\right) - 1 \label{Delta} \\
&=& \frac{\alpha_s(\mu)}{\pi} + {\rm ~higher~orders.}
\end{eqnarray}

\section{Results and systematics}

\begin{figure}[t]
\begin{center}
\includegraphics[height=8cm,trim=28 38 0 20,clip=true]{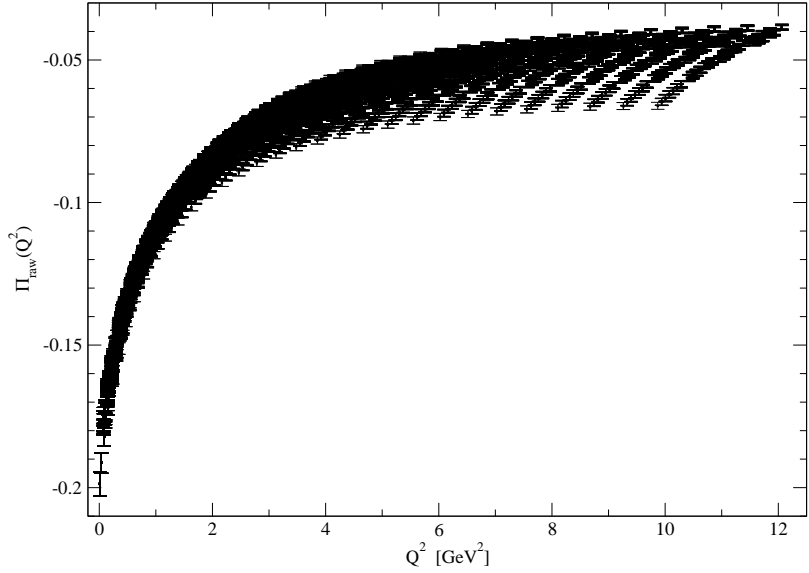}
\end{center}
\vspace{-5mm}
\caption{Raw data for vacuum polarization obtained from 5 configurations for $\bar{s}s$ valence quarks.
Momenta along a single lattice axis are near the bottom of the fishbone pattern.
Momenta along the diagonal of the lattice are near the top of the fishbone pattern.}\label{fig:fishbone}
\end{figure}

\begin{figure}
\begin{center}
\includegraphics[height=8cm,trim=23 38 80 80,clip=true]{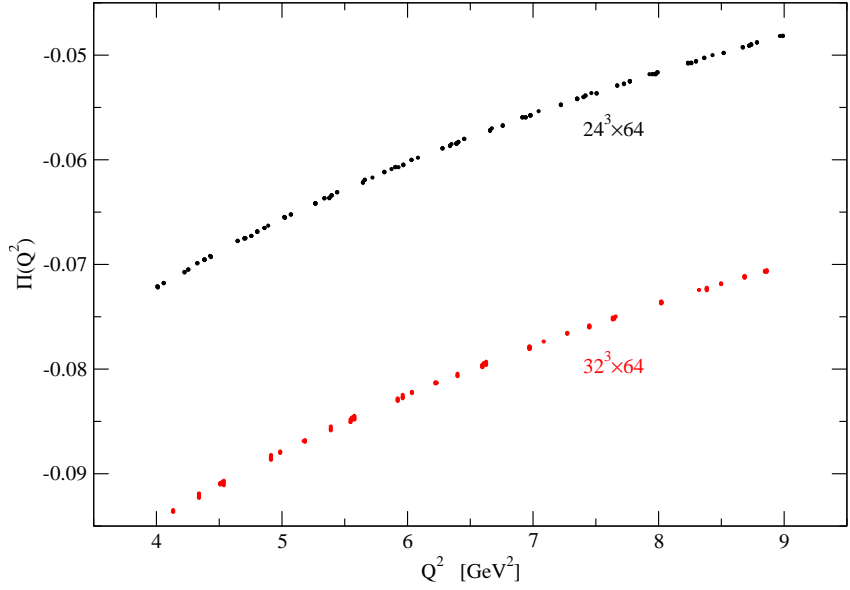}
\end{center}
\vspace{-5mm}
\caption{Data for vacuum polarization after use of Eq.~(\protect\ref{ReflAve}) and an appropriate choice for $|Q_\perp|_{\rm max}$.}\label{fig:ReflAve}
\end{figure}

The raw lattice data obtained from
\begin{equation}
\Pi_{\rm raw}(Q^2) =
\frac{-1}{3\hat Q^2}\left(\delta_{\mu\nu}-\frac{4\hat Q_\mu\hat Q_\nu}{\hat Q^2}\right)\Pi_{\mu\nu}(Q^2)
\end{equation}
show a fishbone pattern due to lattice artifacts, as displayed in Fig.~\ref{fig:fishbone}, making
it difficult to extract physical vacuum polarization as a single-valued function of $Q^2$.
In contrast, the reflection averaging of Eq.~(\ref{ReflAve}) combined with an appropriate choice for $|Q_\perp|_{\rm max}$
produces the smooth curves displayed in Fig.~\ref{fig:ReflAve} that retain a large number of usable $Q^2$ values.

\begin{figure}
\begin{center}
\includegraphics[width=13cm,trim=40 150 40 50,clip=true]{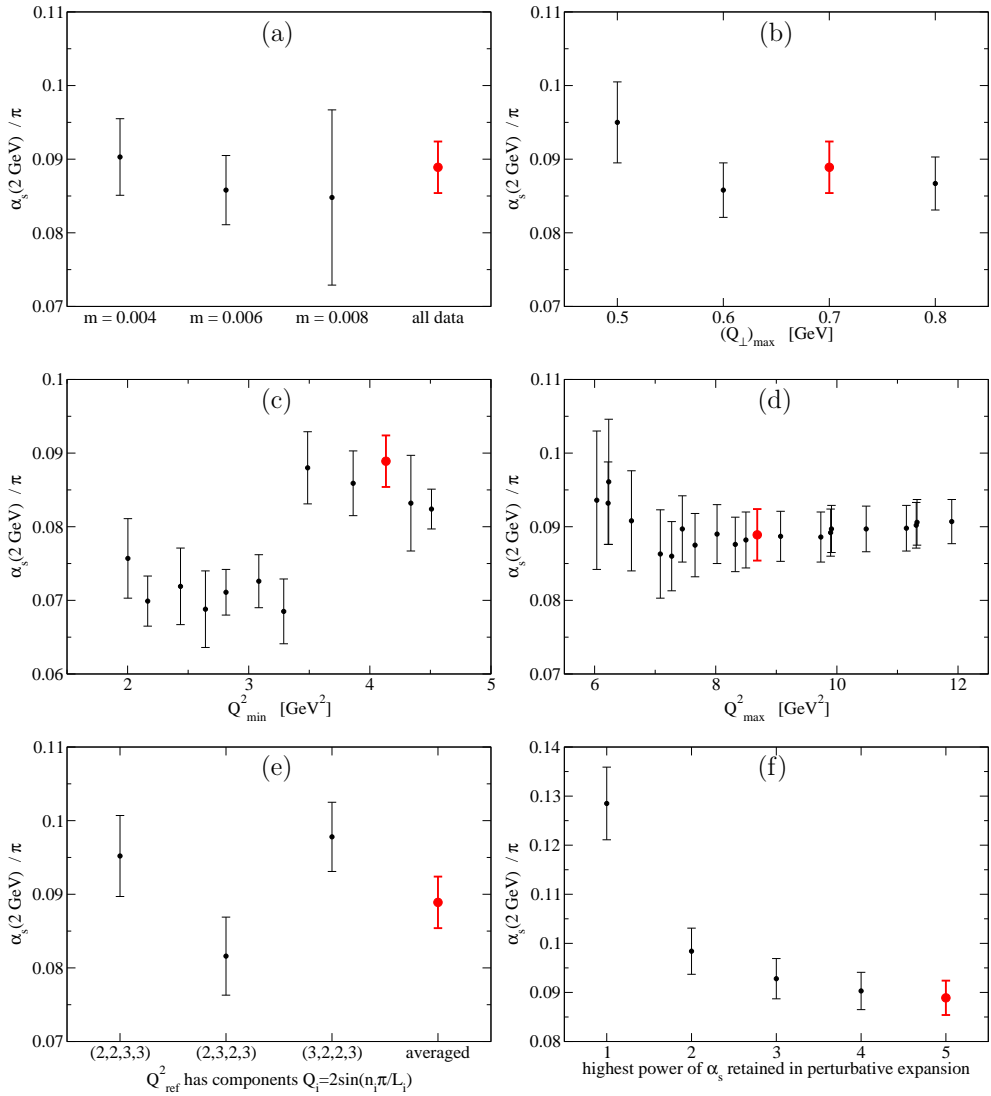}
\end{center}
\vspace{-5mm}
\caption{The strong coupling obtained from the $32^3\times64$ ensemble
(in red), and its dependence on 6 systematics as explained in the
text.}\label{fig:alphas}
\end{figure}

For the $32^3\times64$ ensemble, which has the smaller lattice spacing,
good fits are obtained with just 2 fit parameters.  The fit function is
Eq.~(\ref{fitfunction}) with parameters $\alpha_s(\mu)$ and $c_1$.
All other $c_i$ are set to zero because $c_1$ is sufficient to represent
all of the lattice artifacts in this reflection-averaged data set.
Figure~\ref{fig:alphas} shows the resulting value for $\alpha_s(\mu)$ as
a single red data point that appears on 6 different panels.
Panel (a) indicates that the result is not sensitive to changes in the light
quark mass.
Panel (b) shows insensitivity to our choice of $|Q_\perp|_{\rm max}$; the error
bar grows if $|Q_\perp|_{\rm max}$ becomes too small because an agressive
cut leaves too few data points in the fit.

Experience from $\tau$ decay phenomenology says $Q^2_{\rm min}\gtrsim4$ GeV$^2$
is required to avoid the OPE dangers described above, and panel (c) shows
that these lattice results are consistent with that expectation.
Results should be insensitive to $Q^2_{\rm max}$ if lattice artifacts are
under control, and panel (d) verifies this for our data.

Equation~(\ref{Delta}) requires a subtraction point $Q^2_{\rm ref}$,
and panel (e) shows the statistical variation obtained from choosing any
single direction for $Q_{\rm ref}$ instead of averaging over all equivalent
options.

Panel (f) displays the systematic shift in $\alpha_s$ that would come from
truncating the perturbative expansion at a lower order.
Although the red point does receive input from the estimated coefficient
$c_{50}^A$, our result is insensitive to its precise value.

\begin{figure}
\begin{center}
\includegraphics[width=7cm,trim=0 25 80 80,clip=true]{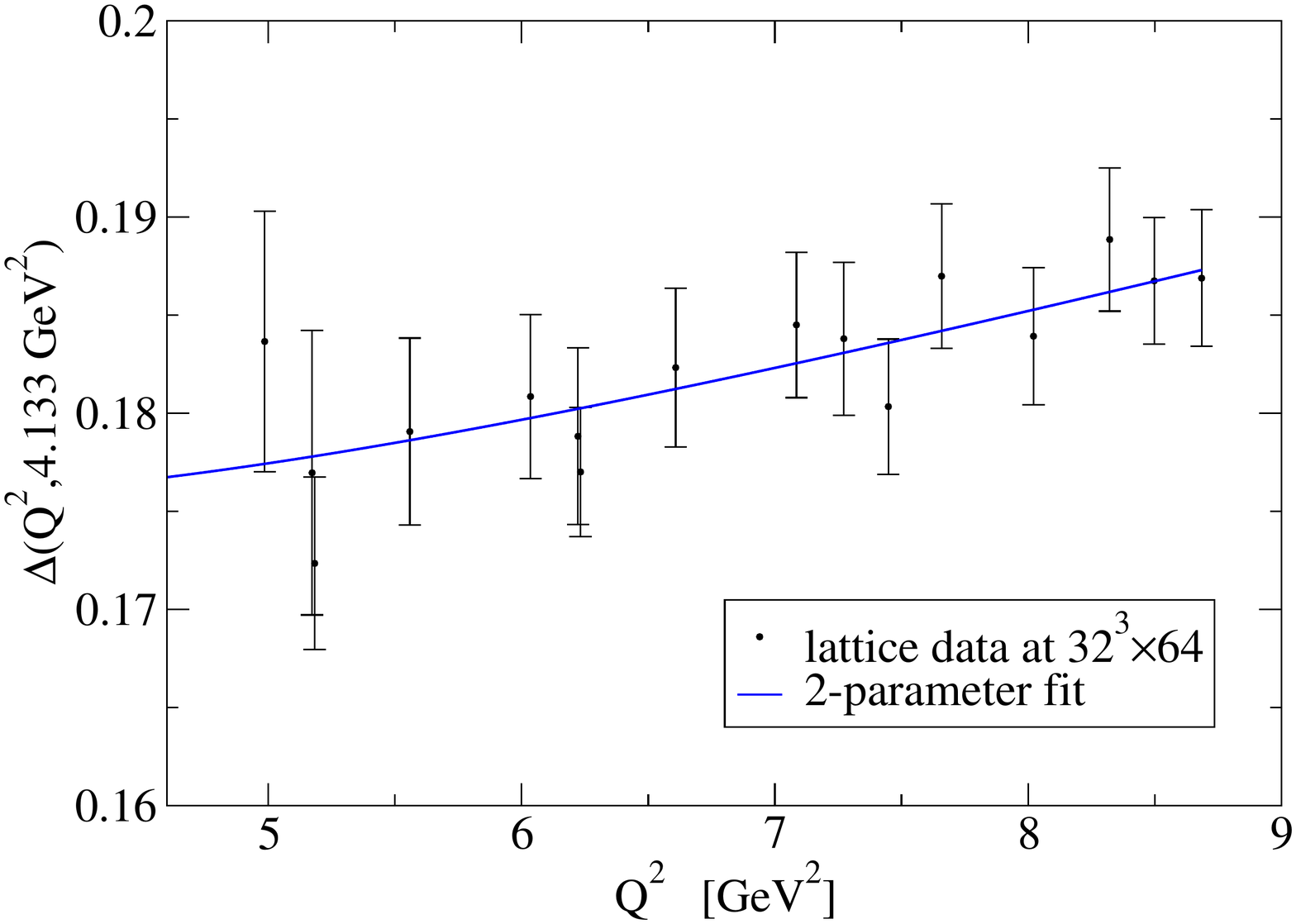}
\includegraphics[width=7cm,trim=0 25 80 80,clip=true]{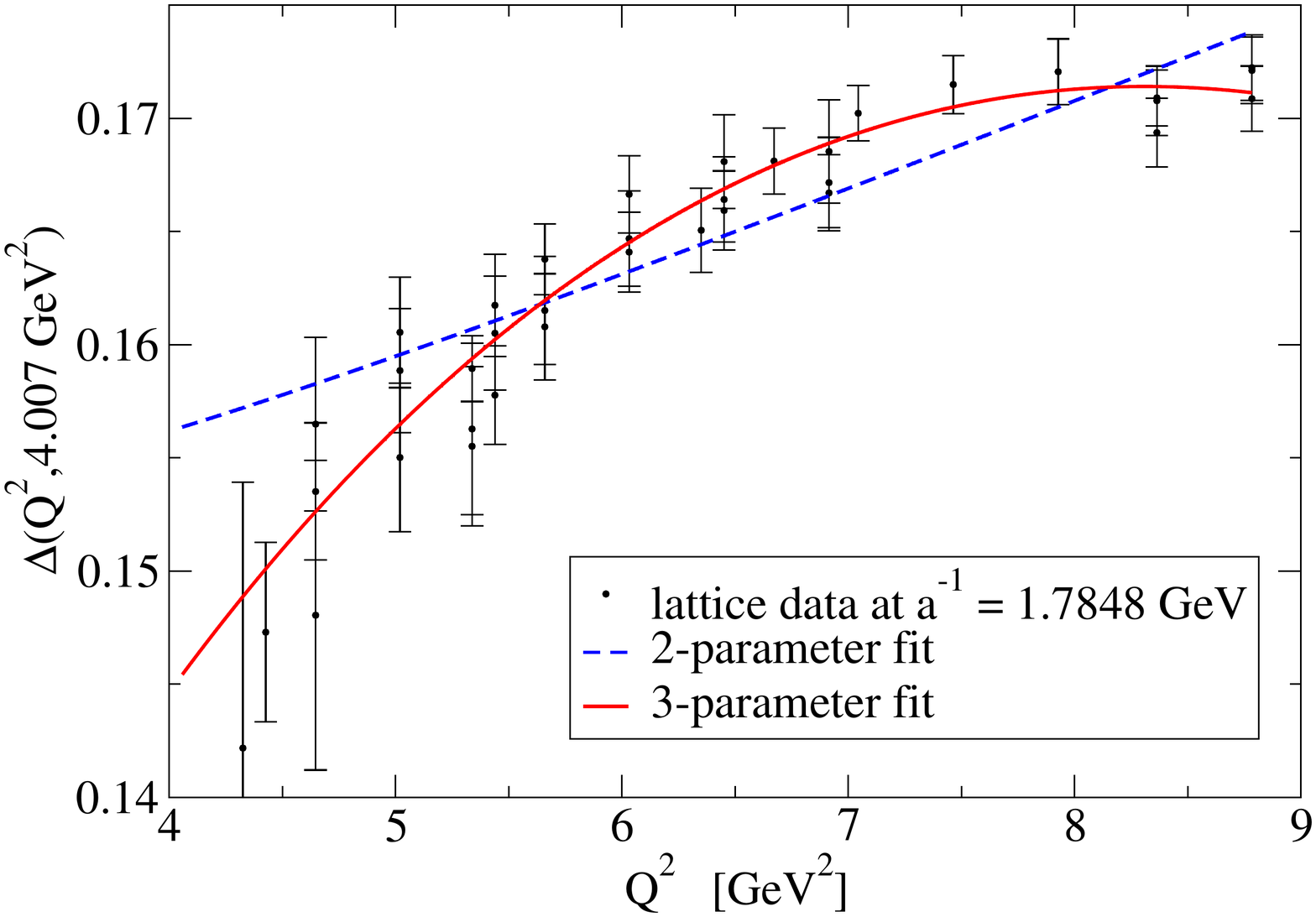}
\end{center}
\vspace{-5mm}
\caption{The left panel shows a useful 2-parameter fit to the finer ensemble.
The right panel shows that the coarser ensemble requires a third parameter in the fit.}\label{fig:finelattice}
\end{figure}

Results displayed in Fig.~\ref{fig:alphas} are for the 32$^3\times$64 ensemble with $a^{-1}=2.38$ GeV.
We should perform a similar analysis for 24$^3\times$64 with $a^{-1}=1.78$ GeV but
two issues arise:
(1) a two-parameter fit (Eq.~(\ref{fitfunction}) with $\alpha_s$ and $c_1$) does not describe the data well, and
(2) a three-parameter fit (Eq.~(\ref{fitfunction}) with $\alpha_s$, $c_1$ and $c_2$) allows a huge error bar for $\alpha_s$.
The situation is displayed graphically in Fig.~\ref{fig:finelattice}.
Unfortunately, we must conclude that a two-parameter fit is not sufficient for the coarse lattice, and that
the statistical precision is not presently available to get $\alpha_s$ from the 3-parameter fit.

\section{Numerical result for $\alpha_s$}

The analysis reported here gives
\begin{equation}\label{ouralphas}
\frac{\alpha_s(2\,{\rm GeV})}{\pi} = 0.0889\pm0.0035
\end{equation}
and running to the $\tau$ mass gives
\begin{equation}
\alpha_s^{(3)}(m_\tau) = 0.296\pm0.013
\end{equation}
which is in excellent agreement with results that use $\tau$ decay data from experiment for
an alternate analysis of the same $\Pi(Q^2)$.
In particular, the recent continuum FESR analysis of the
2013/14 corrected and updated ALEPH hadronic $\tau$ decay data arrived at \cite{Boito:2014yfa}
\begin{equation}
\alpha_s^{(3)}(m_\tau) = \left\{\begin{tabular}{ll} 0.296$\pm$0.010 & {\rm [fixed\mbox{-}order~perturbation~theory],} \\
 0.310$\pm$0.014 & {\rm [contour\mbox{-}improved~perturbation~theory].} \end{tabular}\right.
\end{equation}
Running our result in Eq.~(\ref{ouralphas}) to $m_Z$ in the 5-flavor theory gives
$\alpha_s^{(5)}(m_Z) = 0.1155\pm0.0018$.
For comparison, the FLAG Working Group result is \cite{Aoki:2013ldr}
$\alpha_s^{(5)}(m_Z) = 0.1184\pm0.0012$.

\section{Summary}

We have presented a new implementation to obtain $\alpha_s$ from vacuum polarization at short distances.
It avoids the danger of systematic errors that could arise in low-scale fits
as a result of alternating-sign higher-dimension OPE contributions.
Our method needs $\Pi(Q^2)$ for many off-axis lattice directions.
Our numerical results are competitive with the determination of $\alpha_s$ from $\tau$ decay.
In the future, we hope to use data from finer lattices since this will allow a study of the continuum limit.
\vspace{5mm}

\noindent
{\bf\large Acknowledgments:}
This work was supported in part by NSERC of Canada.


\begin{thebibliography}{99}
\bibitem{Aoki:2013ldr} 
  S.~Aoki {\it et al.},
  Eur.\ Phys.\ J.\ C {\bf 74}, 2890 (2014)
  [arXiv:1310.8555 [hep-lat]].
\bibitem{Shintani:2008ga} 
  E.~Shintani {\it et al.} [JLQCD and TWQCD Collaborations],
  Phys.\ Rev.\ D {\bf 79}, 074510 (2009)
  [arXiv:0807.0556 [hep-lat]].
\bibitem{Shintani:2010ph} 
  E.~Shintani, S.~Aoki, H.~Fukaya, S.~Hashimoto, T.~Kaneko, T.~Onogi and N.~Yamada,
  Phys.\ Rev.\ D {\bf 82}, 074505 (2010)
  [Phys.\ Rev.\ D {\bf 89}, 099903 (2014)]
  [arXiv:1002.0371 [hep-lat]].
\bibitem{Shintani:2014rta} 
  E.~Shintani, H.~J.~Kim, T.~Blum and T.~Izubuchi,
  PoS LATTICE {\bf 2013}, 487 (2014).
\bibitem{Herdoiza:2014jta} 
  G.~Herdoiza, H.~Horch, B.~J\"ager and H.~Wittig,
  PoS LATTICE {\bf 2013}, 444 (2014).
\bibitem{Francis:2014yga} 
  A.~Francis, G.~Herdo\'iza, H.~Horch, B.~J\"ager, H.~B.~Meyer and H.~Wittig,
  PoS LATTICE {\bf 2014}, 163 (2014)
  [arXiv:1412.6934 [hep-lat]].
\bibitem{Boito:2014sta} 
  D.~Boito, M.~Golterman, K.~Maltman, J.~Osborne and S.~Peris,
  Phys.\ Rev.\ D {\bf 91}, no. 3, 034003 (2015)
  [arXiv:1410.3528 [hep-ph]].
\bibitem{Boito:2014yfa} 
  D.~Boito, M.~Golterman, K.~Maltman, J.~Osborne and S.~Peris,
  Nucl.\ Part.\ Phys.\ Proc.\  {\bf 260}, 134 (2015)
  [arXiv:1410.8415 [hep-ph]].
\bibitem{Baikov:2008jh} 
  P.~A.~Baikov, K.~G.~Chetyrkin and J.~H.~Kuhn,
  Phys.\ Rev.\ Lett.\  {\bf 101}, 012002 (2008)
  [arXiv:0801.1821 [hep-ph]].
\bibitem{Arthur:2012yc} 
  R.~Arthur {\it et al.} [RBC and UKQCD Collaborations],
  Phys.\ Rev.\ D {\bf 87}, 094514 (2013)
  [arXiv:1208.4412 [hep-lat]].
\end{thebibliography}
\end{document}